\documentclass[11pt]{article}
\usepackage{fullpage}
\usepackage[round]{natbib}

\usepackage{amsmath,amssymb}
\usepackage{bm}
\usepackage{amssymb}
\usepackage{mathrsfs}
\usepackage{graphicx}
\usepackage{subcaption}
\usepackage{pdfpages}
\usepackage{afterpage}
\usepackage{caption}
\usepackage{multirow}
\usepackage{rotating}
\usepackage[flushleft]{threeparttable}
\usepackage{multirow}
\usepackage{enumitem} 
\usepackage{algorithm2e}
\usepackage{svg}
\usepackage{stackengine}
\usepackage{float}
\RestyleAlgo{ruled}
\SetAlgorithmName{Procedure}{procedure}{List of Procedure}
\usepackage{tikz} 
\usetikzlibrary{matrix}

\usepackage{preamble/star}
\usepackage[group-separator={,},group-minimum-digits={3}]{siunitx}

\usepackage[colorlinks,linkcolor=blue,anchorcolor=blue,citecolor=blue,urlcolor=black]{hyperref}
\usepackage{cleveref}

\newcolumntype{P}[1]{>{\centering\arraybackslash}p{#1}}

\def\$#1\${\begin{align*}#1\end{align*}}

\newcommand{\Rom}[1]{\text{\uppercase\expandafter{\romannumeral #1\relax}}}

\usepackage{color}

\title{Rejoinder: The ICML 2023 Ranking Experiment: Examining Author Self-Assessment in ML/AI Peer Review}
\author{Buxin Su\thanks{University of Pennsylvania.} \and Jiayao Zhang\footnotemark[1] \and Natalie Collina\footnotemark[1] \and Yuling Yan\thanks{University of Wisconsin--Madison.}
\and Didong Li\thanks{Associate Chair of ICML 2023. University of North Carolina at Chapel Hill.} \and Kyunghyun Cho\thanks{Program Chair of ICML 2023. New York University.} \and Jianqing Fan\thanks{Princeton University.} \and Aaron Roth\footnotemark[1]
\and Weijie Su\footnotemark[1]}

\date{May 7, 2026}

\begin{document}
\maketitle

We thank Editor Hongtu Zhu for selecting our article ``The ICML 2023 Ranking Experiment: Examining Author Self-Assessment in ML/AI Peer Review'' \citep{su2025icml} for discussion, and we are grateful for the insightful feedback and thoughtful comments offered by Professor Yoshua Bengio and Dr.\ Dinghuai Zhang, Professor Sanmi Koyejo and Dr.\ Andreas Haupt, Professor Xiao-Li Meng, Professors Weichen Wang and Chengchun Shi, Ying Wang and Professor Mengye Ren, and Professors Linjun Zhang and Lexin Li.

We are delighted to see that many discussants endorse our formulation of machine learning (ML) and artificial intelligence (AI) conference peer review as a statistical problem, provide constructive comments on the Isotonic Mechanism, propose additional signals that can be used in conjunction with the method, and articulate a broader agenda for the reform of peer review in the AI era. Our rejoinder is organized into four themes in the sections that follow.

\section{Peer Review as an Estimation Problem}
\label{sec:why-mechanism-why}

A somewhat radical question that has begun to circulate in the ML/AI community is whether the field could simply abandon peer review and let the ``market'' (that is, readers, gauged through GitHub stars and social media engagement) decide. One motivation for this sentiment is that AI progresses at an unprecedented speed, and much of the research conducted in this field, especially that produced by frontier AI labs such as OpenAI, Anthropic, and DeepSeek, is disseminated as research reports that cannot wait for the peer review process. Another is that the number of submissions to leading ML/AI conferences such as ICML and NeurIPS has grown exponentially, in part because AI is now widely used at various stages of research and substantially accelerates research output, while the supply of
qualified reviewers cannot keep pace with this explosive growth \citep{kim2025position}. For example, ICML 2026 received 26{,}600 valid submissions, exceeding the count at NeurIPS 2025 and thereby becoming, at present, the largest ML/AI conference in history.\footnote{NeurIPS 2026 will receive even more submissions; the point we wish to make is that, for the first time in history, the growth in submissions over a span of half a year is sufficient to offset the gap in submission volume between the two conferences.}

Research papers, however, vary enormously in quality, correctness, novelty, and long-term value, and an average reader cannot reliably evaluate every new submission without some form of filtering. Peer review is, in essence, an attempt to \textit{estimate} the quality of submitted work, a responsibility that academia has upheld throughout its long history. That need has not changed; what has changed is that the estimation problem has become considerably harder to carry out, as more papers must be reviewed at greater speed. Consequently, we see no compelling reason for the scientific community to abandon formal peer review.

For statisticians, viewing peer review as an estimation problem is entirely natural. In ML/AI conferences, each submission typically receives three to five review scores, often on a scale such as 1--8 or 1--10, and the average of these scores is displayed prominently on OpenReview.\footnote{Journal review can also be rendered numerical by encoding decisions such as reject, reject with resubmit, major revision, minor revision, and accept as numerical values.} The mean review score anchors attention and inevitably shapes downstream decisions. Let $\theta_i$ denote the latent quality of submission $i$, where $i$ ranges over tens of thousands of submissions. The displayed average on OpenReview is therefore a noisy estimator of $\theta_i$. Conferences must accordingly make high-stakes decisions
on tens of thousands of submissions within a short turnaround, drawing on noisy scores, written reviews, reviewer confidence ratings, area chair (AC) summaries, and author rebuttals. In this sense, peer review is an estimation problem before it is a decision problem.

This estimation viewpoint is a perspective shared by many of the discussants. Professors Wang and Shi formulate review scores as noisy measurements affected by reviewer bias and variance, and propose reviewer rankings as an additional source of information for estimating latent paper quality. Professors Zhang and Li similarly emphasize latent-quality models, pairwise comparisons, Bayesian or likelihood-based calibration, and reviewer matching under uncertainty. Professor Bengio and Dr.\ Zhang describe the problem as one of finding signal in noise, while Wang and Professor Ren characterize the current reviewer-centric process as a high-variance gatekeeper. Professor Meng situates the issue within the broader context of data science, experimental design, and causal inference. Ultimately, we must rely on statistical studies if we wish to understand deeply how authors, reviewers, editors, and AI tools interact within the peer review system.

\section{The Need for Mechanism Design in Estimation}
\label{sec:need-mech-design}

The framing of peer review as a statistical problem is not a mere formality. It is the starting point for asking what statisticians can contribute to ML/AI peer review, and it indeed served as the motivation for our Isotonic Mechanism. As Wang and Professor Ren put it, the mechanism can be thought of as an approach to eliciting signals from authors in order to denoise review scores. At a high level, this differs from prior work that has focused exclusively on the reviewer side, in that it shifts attention to a complementary actor in the process, namely the authors.

While the intuition that authors possess substantial private knowledge about their submissions is appealing, authors are also stakeholders in the process and may therefore not provide the credible information we seek. This reality calls for mechanism design, the study of how strategic stakeholders can be induced, under appropriately chosen rules, to produce desirable outcomes. Specifically, the question reduces to identifying \textit{constraints} under which authors will truthfully report their private information, and to determining how this information can be incorporated into a statistical procedure for improved estimation of submission quality. This is, in a nutshell, the conceptual origin of the Isotonic Mechanism developed five years ago \citep{su2021you,su2022truthful}.

In fact, there is a long history of implicit mechanism design within the peer review process \citep{aziz2016strategyproof}. Conflict-of-interest rules and randomization of reviewer assignments \citep{xu2023one} are among the many constraints designed to prevent reviewer manipulation, ensuring the collection of credible, albeit still noisy, assessments of submission quality. A notable recent example is the insertion of prompt watermarks in ICML 2026 submissions, aimed at restricting reviewers from utilizing AI tools in specific unauthorized scenarios. However, it is important to note that historically, these rules have been almost exclusively imposed on the reviewers. The Isotonic Mechanism should be understood as a continuation of this tradition, but charting a new direction by designing incentive-aligned rules for the \textit{authors}. 

While many discussants appreciate the novelty of this mechanism, they also raised several practical considerations regarding its real-world deployment, which we address below.

\paragraph{Heterogeneous improvements.} A common concern across the discussions is that the improvement delivered by the Isotonic Mechanism may be heterogeneous across authors. Wang and Professor Ren note that, while the mechanism improves review scores by calibrating authors' private information, it may yield disproportionately larger gains for authors who submit many papers, since a longer ranking is more informative for calibration. Professor Koyejo and Dr.\ Haupt similarly emphasize that variance reduction is more substantial for authors with many submissions than for those with only two or three, raising equity concerns for junior scholars and researchers at smaller institutions.

We begin by emphasizing that the improvement in estimation accuracy is appreciable for rankings of any length. Nevertheless, we agree that this heterogeneity should be taken seriously. At the same time, it is important to distinguish three quantities that are sometimes conflated: the precision of the calibrated score, the probability that a paper receives additional reviewer attention, and the probability that a paper is accepted. Our proposed deployment uses isotonic scores and residuals primarily to direct AC attention, not to make accept/reject decisions automatically. Consequently, the dependence between the length of an author's ranking and the final acceptance decision is, at most, indirect. A more accurate calibrated score arising from a long ranking does not, by itself, translate into a clear advantage in either direction.

Moreover, we argue that the heterogeneity issue may be less severe in practice than it first appears. Our analysis of the ICML 2023 ranking data suggests that the observed heterogeneity may be amplified by the voluntary response pattern in the experiment, and may therefore overstate what would occur under broader participation. The reason is that, for each paper, its modified score is averaged over the calibrated scores produced by each of its authors. This averaging effect tends to smooth out, and substantially reduce, the influence of ranking length across papers. In particular, the influence of authors with very large numbers of submissions would be diluted. This effect is especially pronounced because authors who provide long rankings are typically highly productive senior researchers whose papers tend to have many coauthors.

\begin{figure}[!t]
    \centering
    \begin{subfigure}[b]{0.33\textwidth}
        \includegraphics[width=\textwidth]{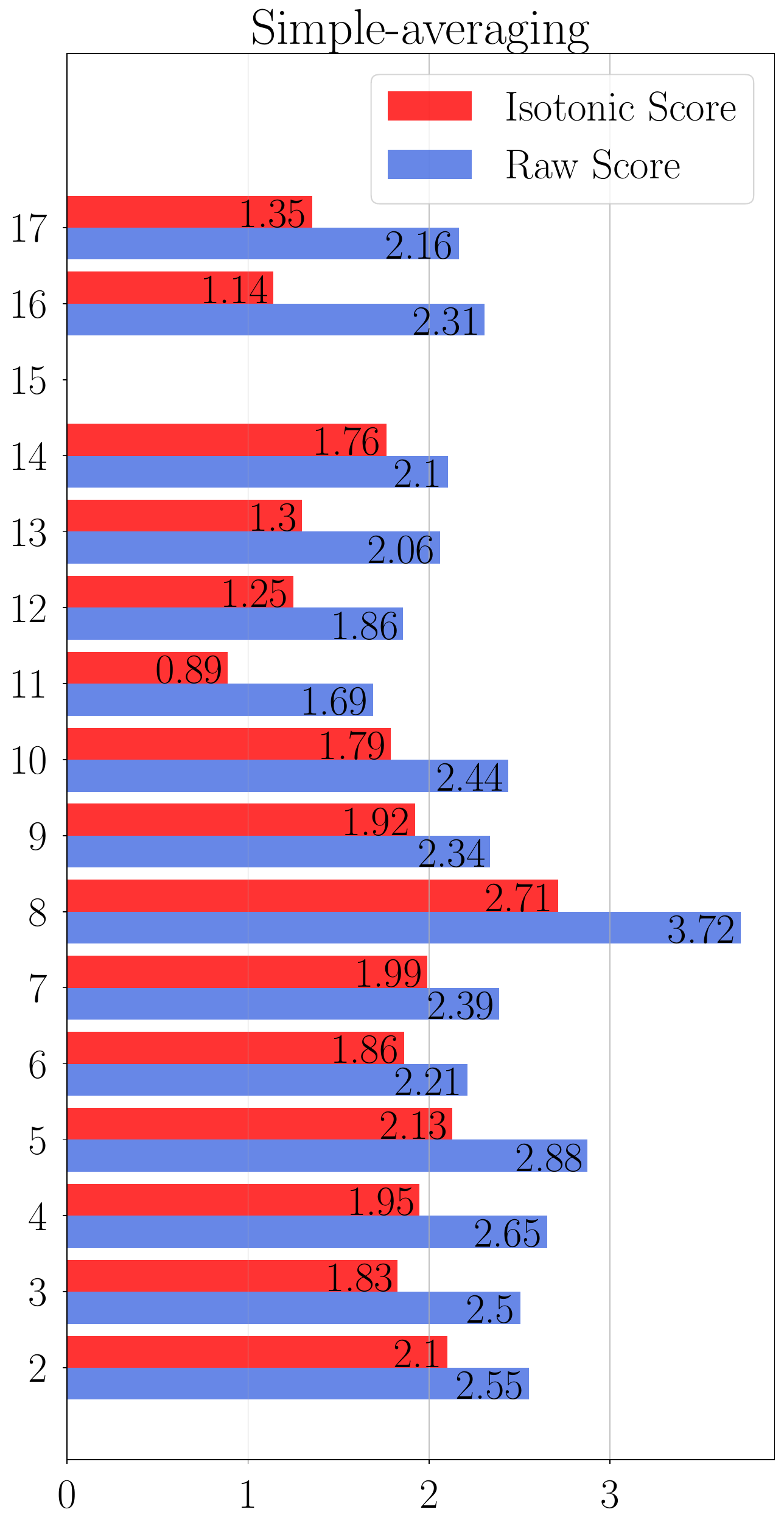}
        \caption{MSE}
    \end{subfigure}
    \hspace{2mm}
    \begin{subfigure}[b]{0.33\textwidth}
        \includegraphics[width=\textwidth]{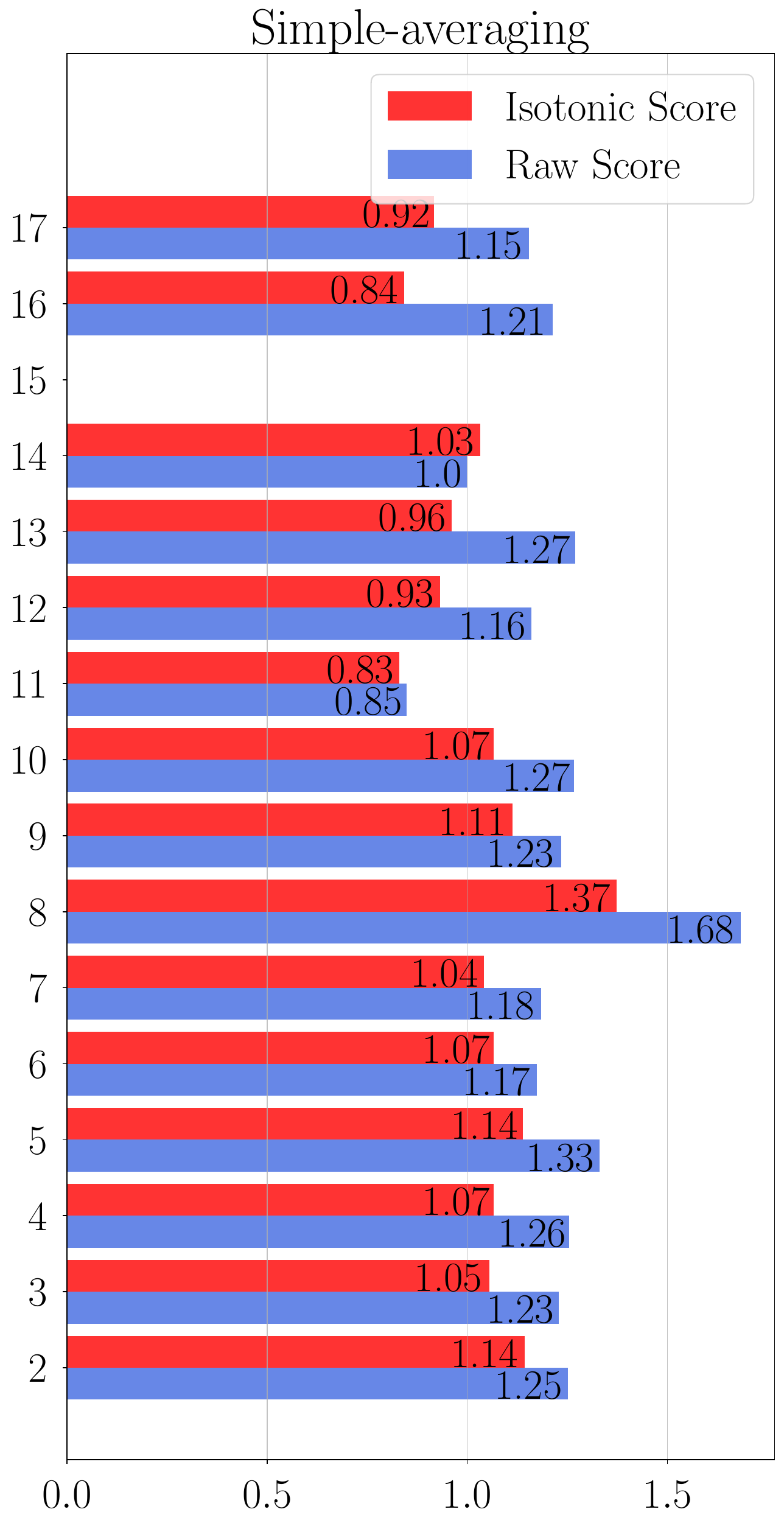}
        \caption{MAE}
    \end{subfigure}
    \caption{MSE and MAE averaged over ICML 2023 authors who submitted rankings of the same length in our survey experiment, using the topic-based partition variant of the Isotonic Mechanism. Overall, relative to the original Isotonic Mechanism, the gains from the modified scores are distributed more evenly across authors with different numbers of submissions. To quantify this pattern, we fit a simple linear regression of percentage improvement on the number of submissions for both versions of the Isotonic Mechanism. For MSE, the slope drops from 1.97 (with p-value $p=0.01$) under the original variant to 1.24 ($p=0.12$) under the topic-based partition variant; for MAE, it correspondingly drops from 1.01 ($p=0.05$) to 0.63 ($p=0.29$). The reported $p$-values test the null hypothesis that the regression slope is zero.}
    \label{fig:unbiased_improve}
\end{figure}

Stepping back, if the goal is to explicitly minimize this potential heterogeneity, a simple mitigating approach is to partition a prolific author's long list of submissions into smaller, topic-based blocks, and then apply the Isotonic Mechanism independently within each block \citep{wu2023isotonic}. Figure~\ref{fig:unbiased_improve} reports the mean squared error (MSE) and mean absolute error (MAE) under this strategy. For the subgroup with the longest ranking length (17, comprising only two authors), the calibrated-score MSE rises from 1.00 to 1.35 under the new variant, compared with 2.16 for raw review scores; for authors with only two submissions, the MSE decreases from 2.17 to 2.10, compared with 2.55 for raw review scores. More importantly, the relative improvement delivered by the modified scores becomes considerably more homogeneous across authors with different numbers of submissions, as quantified by the regression slopes reported in the figure caption. Even with this equalizing adjustment, the simple strategy still achieves an approximately 21\% reduction in MSE. The point we wish to make is that fairness concerns can be encoded directly into the mechanism, rather than treated solely as a post hoc diagnostic.

To establish higher statistical validity, a randomized experiment quantifying the effect of ranking length would yield more convincing and quantitative conclusions. Such an experiment is currently underway at ICML 2026, where the first and last authors of the present paper are serving as Associate Chair and Integrity Chair, respectively, and are incorporating calibrated scores into the decision-making process. We expect to report findings later this year.

In a related observation, Professors Wang and Shi point out that authors with only a single submission cannot directly participate in the ranking mechanism, and may therefore face noisier evaluations. While this is a valid point at the individual author level since only 24.3\% of authors had at least two submissions at ICML 2023, it is substantially mitigated at the paper level. In fact, 77.0\% of all submitted papers feature at least one coauthor who has multiple submissions. Consequently, the Isotonic Mechanism could cover 77.0\% of the total submissions.

\paragraph{Strategic behavior.} Several discussants raise the concern that authors may behave differently in a deployed setting, where calibrated scores influence acceptance decisions and paper awards \citep{wen2026recommending}, than they did in our ICML 2023 experiment, in which the submitted rankings carried no consequences.

This is a valid concern, fundamentally because the incentives in real deployment can be more complex than the utility model assumed in the development of the Isotonic Mechanism \citep{su2021you,su2022truthful,yan2023isotonic}. For example, authors may favor papers on which they are the first author, as Wang and Professor Ren note, or they may favor papers by graduating students or junior coauthors, as observed by Professor Koyejo and Dr.\ Haupt, Professor Bengio and Dr.\ Zhang, and Professor Meng. A principal investigator who ranks many lab papers may thus, in effect, convert part of the conference review process into an internal lab-allocation problem.

Threshold-based decision rules, which accept a paper once its average score exceeds a threshold, create another strategic channel, although such rules are not commonly applied in their pure form. As Professor Koyejo and Dr.\ Haupt and Professor Bengio and Dr.\ Zhang point out, an author may seek to maximize the number of accepted papers rather than report the true quality ordering. In such a setting, a paper that appears safely above the acceptance threshold might be ranked lower so that a borderline paper can be ranked higher. Professor Meng describes the related possibility that a particularly strong paper could subsidize weaker sibling submissions. More generally, authors may intentionally elevate weaker papers above stronger ones when doing so serves career, lab, or acceptance-count objectives. This concern is also connected to the earlier observation that longer rankings yield greater calibration benefits: if long rankings are valuable, authors may have incentives to submit more papers, coordinate rankings through senior lab members, or, as Wang and Professor Ren note in the context of generative AI, even generate low-effort additional submissions in order to alter the ranking structure.

The fact that major ML/AI conferences cycle continuously throughout the year makes these concerns more pressing. Professors Wang and Shi emphasize that, because conferences recur annually, authors can learn over time how a deployed mechanism responds. Strategic behavior need not be perfectly planned in the first year; it can emerge through multi-year experimentation, imitation, and adversarial gaming. Optional participation introduces a further selection problem, since authors who choose to submit rankings may differ systematically from those who do not. Mandatory rankings, randomized encouragement, or staggered deployment can help diagnose this selection, but they do not eliminate the need to monitor behavior once rankings carry consequences.

We view these comments as a constructive agenda. First, the incentive landscape depends strongly on the use case. Using isotonic residuals to flag cases for additional AC examination creates weaker and less predictable incentives than using calibrated scores directly for acceptance. Award selection, emergency-reviewer recruitment, and review-quality auditing each present distinct strategic profiles. Second, strategyproofness may need to be made psychologically and institutionally transparent. If authors do not understand why truthfulness is in their interest, or if they optimize for an objective different from the utility assumed in the Isotonic Mechanism, the mechanism may fail in practice even when the formal result remains correct. Third, the same lesson applies beyond author rankings. Reviewer recognition systems can encourage reviewers to optimize for speed or public reputation rather than review quality, and post-publication or community-review systems can be manipulated through popularity, networks, or coordinated promotion. Low-stakes uses, with randomized or staggered deployment, are therefore not merely a matter of convenience but a necessary starting point for understanding how strategic behavior emerges when theoretical and practical incentives diverge. Such understanding will not be credible without experiments in which rankings are consequential. ICML 2026 represents a first example of such a deployment, and, as Professor Koyejo and Dr.\ Haupt note, more will be needed in the future.

While many of these strategic behaviors can be mitigated through continuous monitoring and ad hoc filtering, a more principled approach is to formally design new mechanisms that explicitly incorporate these complex incentive structures to ensure socially optimal outcomes for peer review. Crucially, mechanism design in this space cannot be decoupled from statistical modeling, as the ultimate objective remains the accurate estimation of underlying submission quality. This intersection presents a rich and expansive frontier of research opportunities for statisticians.

\section{Other Signals and Extensions}
\label{sec:meth-extens}

Statistical methodology can improve peer review along multiple dimensions, certainly extending well beyond the use of authors' own rankings. Indeed, several discussants propose signals beyond the author ranking used in our experiment. These signals can be combined with author rankings to further enhance the quality of peer review.

Professor Bengio and Dr.\ Zhang suggest asking authors to evaluate their submissions along multiple axes, such as novelty, rigor, and predicted long-term impact. We find this direction appealing because a single scalar review score may not capture all dimensions of scientific value. A paper may be technically imperfect yet unusually creative; another may be rigorous but incremental. If conference organizers wish to assemble a portfolio of accepted papers that balances reliability with long-term impact, a one-dimensional score will not suffice. Even within our existing framework, although authors are asked to rank submissions by perceived overall quality, we can additionally solicit, for instance, which papers fall in the top 30\% or 50\% under a long-term-impact criterion. As an aside, a follow-up study of the ICML 2023 ranking experiment shows that authors' own rankings provide a powerful predictor of how many citations their papers will receive \citep{su2025find}.

Professors Wang and Shi, as well as Professor Bengio and Dr.\ Zhang, propose incorporating reviewer rankings, which are a natural complement to author rankings. Reviewers already compare papers implicitly when they assign scores, and a desirable property of reviewer rankings is that they are largely invariant to reviewer bias. Professors Wang and Shi develop a method based on reviewer rankings that yields strong estimates of submission quality even when reviewer bias is substantial. We view reviewer rankings and author rankings as complementary signals: reviewer rankings help calibrate score scales across reviewers, while author rankings provide within-author comparative information that reviewers do not observe. An interesting research direction is the joint integration of rankings from both authors and reviewers. Professors Wang and Shi report promising preliminary results from combining the two, and further effort can be devoted to developing adaptive weighting schemes for the two types of rankings.

Professors Zhang and Li suggest pairwise comparisons, structured metadata, and self-predictions as alternatives or complements to author rankings. In general, authors may find pairwise comparisons easier to provide than a complete ranking, especially when the list of submissions is long. A useful property of the Isotonic Mechanism is that its truthfulness extends beyond strict rankings to a broad class of pairwise comparison structures, including ``coarse rankings'' in which an author partitions her submissions into several blocks and asserts that all submissions in block $i$ are better than those in block $i+1$. The current ranking portal at OpenReview, for example, already permits ties. An interesting question for future research is how to adaptively design pairwise comparisons for authors that ensure truthfulness while reducing the elicitation burden.

Structured metadata provides a different kind of signal. Authors may know whether a paper is primarily theoretical, empirical, methodological, or interdisciplinary, and which type of reviewer expertise is most relevant to its evaluation. Such metadata may be of limited use for directly scoring papers, but highly useful for reviewer matching, AC triage, and the flagging of reviewer miscalibration or mismatched expertise. Prior work suggests that authors can have reasonably informative expectations about review outcomes \citep{rastogi2022authors}, but self-predictions should be used cautiously, since optimism bias and strategic incentives are unavoidable.

Such structured metadata may also be helpful when extending the mechanism from conferences to journals. Professors Zhang and Li emphasize that this extension is not a matter of simple transplantation. Conference review typically involves fixed deadlines, many simultaneous submissions, and short review cycles, whereas journal review more often involves rolling submissions, fewer simultaneous papers from the same authors, and multiple rounds of revision. We agree that these structural differences matter. In a journal setting, author-side signals might therefore need to be profoundly reimagined, perhaps, as suggested by Professor Bengio and Dr.\ Zhang, by asking authors to position a new submission relative to their previously published works of a similar nature. However, designing a mathematically rigorous, truthful mechanism for this longitudinal setting remains an open and challenging problem. On the other hand, the extended timelines inherent in journal reviews naturally afford the opportunity to collect and process much richer, highly structured metadata.

\section{Toward Human-Centered Peer Review in the AI Age}
\label{sec:syst-level-depl}

From a broader perspective, scientific publication is entering a period of unusual uncertainty. AI tools are increasingly used not only to polish writing, but also to assist with literature search, experimentation, and the early stages of research design. One visible consequence is the rapid growth in submissions to many publication venues, as noted earlier. A deeper consequence is that some traditional cues of paper quality have become less reliable. Polished language, well-structured exposition, and even technically elaborate derivations may no longer provide the same evidence about the underlying scientific contribution when these elements can be substantially produced with the assistance of large language models. Looking further ahead, increasingly automated AI research systems, which may be deployed at scale within a few years, could produce a volume of papers beyond imagination \citep{Down2025AISlop}. These developments do not render peer review obsolete; rather, they make the reform of peer review urgent.

The same shift is occurring on the reviewer side \citep{kim2025position}. The question is no longer whether AI should be used in peer review. In practice, authors, reviewers, ACs, and editors all have access to AI tools, and several venues such as ICML 2026 and AAAI 2026 have already begun to incorporate AI tools into their review processes. The more relevant question is how AI can be used so that peer review becomes more reliable, more transparent, and less burdensome, while preserving human responsibility for scientific judgment. A guiding principle that we find useful is that AI should support bounded and auditable tasks, while humans remain accountable for the final judgment. This principle is not merely conservative. Scientific work is ultimately written for, interpreted by, and built upon by human
communities. Judgments about novelty, validity, significance, taste, ethics, and long-term value cannot be reduced to automated scoring without losing something central to scientific communication.

We therefore view the Isotonic Mechanism as one instance of a broader human-centered principle. It does not replace reviewers with an automatic decision rule; rather, it brings an additional human signal into the process by eliciting authors' comparative judgments and combining that signal conservatively with review scores. As AI becomes more pervasive in both writing and reviewing, such human signals may become more valuable, not less. Authors, reviewers, and ACs possess contextual knowledge that is difficult to reconstruct from text alone. The design challenge is to elicit this knowledge in forms that are useful, incentive-aware, and empirically auditable.

Professor Meng's discussion is especially helpful in this respect, as it reminds us that peer review is a human system before it is a computational system. His Human+AI proposal points toward a productive division of labor: AI can help reviewers and ACs organize information, retrieve relevant context, check consistency, and notice possible omissions, while human participants remain responsible for scientific interpretation and the final recommendation. Echoing Professor Meng's formulation, we endorse the view that AI should raise the floor, not set the bar. AI tools may improve timeliness and completeness, but they should not be treated as independent judges of acceptability. AI-generated reviews can hallucinate, miss subtle technical flaws, leak confidential information when used carelessly, and create a false sense of objectivity \citep{yuan2022can,liang2023can}.

Human-centered reform also requires better reviewer-side design. Professors Linjun Zhang and Lexin Li discuss reviewer matching, reviewer-quality estimation, reviewer feedback, and reviewer recognition, and we agree that these are central components of any serious reform. Reviewer-side signals such as expertise, calibration, timeliness, helpfulness, and willingness to engage in discussion can be modeled with hierarchical or dynamic methods and used for better assignment, training, and recognition \citep{shah2018design,stelmakh2021novice,goldberg2025peer}. AI may assist this process by summarizing review histories, detecting mismatches between paper topics and reviewer expertise, or flagging unusually sparse or inconsistent reports. The purpose of such tools, however, should be to support better human reviewing, not to replace reviewers or to create brittle leaderboards. Recognition systems should reward careful and constructive judgment, rather than mere speed, volume, or surface-level agreement with later decisions.

For the statistics community in particular, the AI age may also call for new publication formats. A recent large-scale analysis of the literature shows that contemporary statistical research increasingly engages themes at the intersection of statistics and AI, including deep learning theory, interpretability, privacy, and causal ML \citep{he2026emerging}. In retrospect, the lasting scientific influence of statistical methodology is visible well beyond statistics journals. For example, a news feature in \textit{Nature} listed Breiman's random forests among the ten most cited papers of the twenty-first century across the broader scientific community \citep{pearson2025exclusive}.

In light of this, one modest possibility is to experiment with conference proceedings or other selective, citable outlets for shorter papers in emerging directions. This may be especially valuable for work at the interface of AI and other domains, where progress is often iterative and where waiting for a single definitive result may unduly slow communication. Wang and Professor Ren point toward a broader publishing ecosystem in which papers are disseminated more rapidly and then evaluated continuously through author, reviewer, and community signals. We view this as a promising direction, provided that continuing evaluation is itself audited for bias, popularity effects, and strategic manipulation. A realistic goal may be a hybrid system: calibrated initial screening, faster dissemination when appropriate, and post-publication signals that accumulate transparently over time.

In summary, peer review in the AI age remains an estimation problem, but the data-generating process is becoming considerably more complex. AI is changing how papers are written, how submissions are produced, how reviews are drafted, and how readers interpret signals of quality. The challenge, therefore, is not to choose between human judgment and automated assistance. It is to design systems that combine imperfect human and machine-generated signals in principled, transparent, and empirically validated ways. This is a genuine opportunity for the statistics community. Our expertise in statistical modeling is directly relevant to the future of peer review across scientific fields. We hope this exchange will encourage more statisticians to treat peer review itself as an important object of statistical research, and as a setting in which careful methodology can serve the broader scientific community.

\bibliographystyle{apalike}
\bibliography{ref}

\end{document}